\documentclass[doublecol]{epl2} 

\usepackage{amssymb}
\title{Stationary states and fractional dynamics in systems with long range
interactions }

\author{T. L. Van Den Berg\inst{1} \and D. Fanelli\inst{2} \and X. Leoncini\inst{3} }
\shortauthor{T. L. Van Den Berg \etal}

\institute{                    
  \inst{1} IM2NP, Aix-Marseille Université, Centre de
 Saint J\'er\^{o}me, Case 142,  13397 Marseille Cedex 20, France\\\\
  \inst{2} Dipartimento di Energetica Sergio Stecco, Universita
di Firenze, via s. Marta 3, 50139 Firenze, Italia e Centro interdipartimentale
per lo Studio delle Dinamiche Complesse (CSDC)\\\\
\inst{3} Centre de Physique Th\'eorique\thanks{Unit\'e Mixte de Recherche (UMR 6207) du CNRS, et des universit\'es Aix-Marseille
I, Aix-Marseille II et du Sud Toulon-Var. Laboratoire affili\'e à la
FRUMAM (FR 2291).}, Aix-Marseille Universit\'e, CNRS, Luminy, Case 907, F-13288 Marseille cedex 9, France\\}


\abstract{Dynamics of many-body Hamiltonian systems with long range interactions
is studied, in the context of the so called $\alpha-$HMF model. Building
on the analogy with the related mean field model, we construct stationary
states of the $\alpha-$HMF model for which the spatial organization
satisfies a fractional equation. At variance, the microscopic dynamics
turns out to be regular and explicitly known. As a consequence, dynamical
regularity is achieved at the price of strong spatial complexity,
namely a microscopic inhomogeneity which locally displays scale invariance.}

\begin{document}

\maketitle
The out-of-equilibrium dynamics of many body systems subject to long
range couplings defines a fascinating field of investigations, which
can potentially impact different domains of applications \cite{Campa09}. In a long
range system every constituent is simultaneously solicited by the
ensemble of microscopic actors, resulting in a complex dynamical picture.
This latter scenario applies to a vast realm of fundamental problems,
including gravity and plasma physics, and also extends to a rich variety
of cross disciplinary studies \cite{Dauxois_book2002}.
In particular, and with specific emphasis on the peculiar non equilibrium
features, long range systems often display a slow relaxation to equilibrium.
They are in fact trapped in long-lasting out of equilibrium regimes,
termed in the literature Quasi Stationary States (QSS) which bear
distinct characteristics, as compared to the corresponding deputed
equilibrium configuration. A paradigmatic representative of long range
interactions, sharing the mean field viewpoint, is the so called Hamiltonian
Mean Field model, which describes the coupled evolution of $N$ rotators,
populating the unitary circle and interacting via a cosines like potential.
In the limit of infinite system size the discrete HMF model is described by the Vlasov equation for the evolution of the single particle distribution
function\cite{Chavanis-RuffoCCT07}. This leads to a statistically based
treatment, inspired by
 the seminal work of Lynden-Bell, which revealed
the existence of two different classes of QSS, spatially homogeneous
or inhomogeneous \cite{Chavanis2006,Antoniazzi2007}. More recently, and still with reference
to the HMF case study, stationary states have been constructed using
a dynamical scheme which exploited the formal analogy with a set of
uncoupled pendula \cite{Leoncini09b}. This represented a substantial
leap forward in the understanding of the dynamical properties of the QSS
in the HMF model, beyond the aforementioned statistical approach.
Indeed, it was understood that the microscopic dynamics in the inhomogeneous
stationary state is regular and explicitly known, an observation which
contributed to shed light onto the puzzling abundance of emerging
regular orbits as revealed in \cite{Bachelard08}.

These last results have been though obtained in the framework of mean
field systems: The actual distance between particles does no explicitly
appear in the HMF potential. In this letter we aim at bridging this
gap, by focusing on the so called $\alpha-$HMF model \cite{Anteneodo98}.
We shall here concentrate on the long range version of the model which implies 
assuming $0<\alpha<1$ \cite{Dauxois_book2002}. In this
long range version
 the model
behaves at equilibrium as the HMF, see for instance \cite{Tamarit00,Campa03,Campa09}.
We may then ask if the same correspondence applies to the out of equilibrium
dynamics. QSS exists for the $\alpha-$HMF model,as shown in\cite{Campa02}. However, can one still
appreciate the asymptotic trend towards regularity? How does the spatial
organization impact on the aforementioned features? It was also shown
recently that fractional calculus may be a crucial ingredient when
dealing with long range systems \cite{Zaslav07}, and we shall see
how this point enters the picture in the considered case. To anticipate
our findings, we shall here show that all stationary states of the
HMF model are shared by the $\alpha-$HMF model: Particles still exhibit
regular orbits, at the price of a enhanced microscopic
spatial complexity, which materializes in a small scale inhomogeneity
being locally scale invariant. Let us start by introducing the governing
Hamiltonian which can be cast in the form:

\begin{equation}
H=\sum_{i=1}^{N}\left[\frac{p_{i}^{2}}{2}+\frac{1}{2\tilde{N}}\sum_{j\ne i}^{N}\frac{1-\cos\left(q_{i}-q_{j}\right)}{\Vert i-j\Vert^{\alpha}}\right]\:,\label{eq:Hamiltonian_HMF}\end{equation}
where $q_{i}$ stands for the orientation of the rotor occupying the
lattice position $i,$ while $p_{i}$ labels the conjugate momentum.
The quantity $\Vert i-j\Vert$ denotes the shortest distance on the
circle of circumference $N$. The coupling constant between classical
rotators decays as a power law of the sites distance. The HMF limit
is recovered for $\alpha=0$. For $N$ even, we have \begin{equation}
\tilde{N}=\left(\frac{2}{N}\right)^{\alpha}+2\sum_{i=1}^{N/2-1}\frac{1}{i{}^{\alpha}}\:,\label{eq:Tilde_N}\end{equation}
which guarantees extensivity of the system. The equations of motion
of element $i$ are derived from the above Hamiltonian and can be
written as follows\begin{equation}
\dot{p_{i}}=-\sin(q_{i})C_{i}+\cos(q_{i})S_{i}=M_{i}\sin(q_{i}-\varphi_{i})\:.\label{eq:p_dot_bis}\end{equation}
where use has been made of the following global quantities: \begin{eqnarray}
C_{i} & = & \frac{1}{\tilde{N}}\sum_{j\ne i}\frac{\cos q_{j}}{\Vert i-j\Vert^{\alpha}}\label{eq:C_simple_N_tilde}\\
S_{i} & = & \frac{1}{\tilde{N}}\sum_{j\ne i}\frac{\sin q_{j}}{\Vert i-j\Vert^{\alpha}}\:,.\label{eq:S_simple_N_tilde}\end{eqnarray}
These identify the two components of a non-local magnetization per
site, with modulus $M_{i}=\sqrt{C_{i}^{2}+S_{i}^{2}}$ , and phase
$\varphi_{i}=\arctan(S_{i}/C_{i})$. In doing so, one brings into
evidence the formal analogy with the HMF setting. Indeed we notice
that each individual $\alpha-$HMF particle obeys a dynamical equation
which closely resembles that of a pendulum. This observation represented
the starting point of the analysis carried out in \cite{Leoncini09b},
where stationary states were constructed from first principles. More
specifically, the authors of \cite{Leoncini09b} imagined that the
system of coupled rotators reached a given 
 stationary state, characterized
by a constant magnetization in the limit for $N\rightarrow\infty$
. Then, the HMF model is mapped into a set of $N$ uncoupled pendula,
constrained to collectively return a global magnetization identical
to that driving their individual dynamics. Self-consistency is hence
a crucial ingredient explicitly accommodated for the formulation proposed
in \cite{Leoncini09b}. Technically, the stationary state is built
by exploiting the ergodic measure on the torus originating from the
pendulum motion, a working ansatz that we cannot invoke in the context
of the $\alpha-$HMF, due to site localization. Eventually we will
overcome this difficulty by considering the continuous limit. We notice
that, for large $N$ , and since $0<\alpha<1$, the following estimate
applies

\begin{equation}
\tilde{N}\approx\frac{2}{1-\alpha}(N/2)^{1-\alpha}\:.\label{eq:N_tilde_large_N}\end{equation}
We can then use expression (\ref{eq:N_tilde_large_N}) in Eq.(\ref{eq:C_simple_N_tilde})
and, as $N\rightarrow\infty$ , introduce the continuous variables
$x=i/N$ and $y=j/N$ and arrive to the following Riemann integral 

\begin{equation}
C(x)=\frac{1-\alpha}{2^{\alpha}}\int_{-1/2}^{1/2}\frac{\cos\left(q(y)\right)}{\Vert x-y\Vert^{\alpha}}dy\:,\label{eq:C_of_x}\end{equation}
where $\Vert x-y\Vert$ represents the minimal distance on a circle
of circumference one. By invoking the Riemann-Liouville formalism on the
circle, we recognize the fractional integral $I^{1-\alpha}$.
 To be more precise, the Riemann-Liouville formalism on a circle cannot be defined in a consistent way, 
as the Riemann-Liouville operations cannot map periodic functions into periodic functions. However,  the integration 
is performed over the whole circle in Eq.(\ref{eq:C_of_x}). And, in fact, 
we are relying on fractional integrals of the potential type, which means we are actually considering a linear combination of  Weyl fractional derivatives, see for instance \cite{Samko_Book93} for more details. Consequently we can
write \begin{equation}
C(x)=\frac{1-\alpha}{2^{\alpha}}\Gamma(1-\alpha)I^{1-\alpha}\left(\cos q(x)\right)\:.\label{eq:Fractional_C}\end{equation}
A similar relation holds for the $S(x)$ component. Studying the $\alpha-$HMF
dynamics implies characterizing the evolution of the scalar fields
$q(x,t)$ and $p(x,t)$ which are ruled by the fractional (non-local)
partial differential equations\begin{eqnarray*}
\frac{\partial q}{\partial t} & = & p(x,t)\\
\frac{\partial p}{\partial t} & = & \frac{\mu}{2^{\alpha}}\Gamma(\mu)\left(-\sin(q)I^{\mu}\left(\cos q\right)+\cos(q)I^{\mu}\left(\sin q\right)\right)\:.\end{eqnarray*}
where $\mu=1-\alpha$. 

At variance with the simpler HMF ($\alpha=0$)
model, the spatial organization $q(x)$ matters within the general
setting $0<\alpha<1$, an observation which, as anticipated above, poses
technical problems to a straightforward extension of the analysis
in \cite{Leoncini09b}. 

In order to carry on the study and mimick the mean field situation, 
we turn to considering the particular condition where $C(x)=\langle C\rangle=constant$, where $\langle \cdots \rangle$ denotes a spatial average. At the same time, and without losing generality, 
we require $I^{\mu}\left(\sin q\right)=0$, which corresponds to setting to zero the $S$ component of the magnetization. Hence, we choose to deal with a constant 
magnetization amount $M$ such that $C(x)=M$.
 
Note that the translation invariance
along the lattice is also likely to statistically lead to the configuration here scrutinized \cite{Campa03}.
 For finite size systems, the identity $C(x)=M$ can be solely matched
by the trivial state where all $q$ are equal. Conversely, in the
infinite $N$ limit, 
the latter assumption, or equivalently $dC/dx=0$, implies: \begin{equation}
{\cal D}^{\alpha}\cos q = \frac{d^{\alpha}\cos q}{dx^{\alpha}}=0\:.\label{eq:fractional_Steady_state}\end{equation}
where we recalled expression (\ref{eq:Fractional_C}) and
where the operator $D^{\alpha}$ stands for the fractional derivative.
Trivial states (q(x)=Cte) as evidenced in the finite size approximation are also
solutions of this equation. However, as we will argue in the following,
the continuum limit enables us to compute an independent set of admissible
solutions. This task is accomplished by exploiting the fact that $\alpha<1$,
which makes the integral $\int1/\Vert x\Vert^{\alpha}dx$ convergent
near $0$ and that the function $1/\Vert x-y\Vert^{\alpha}$ is smooth.
Noticing that we rewrite Eq. (\ref{eq:C_of_x}) as: \begin{equation}
C(x)=\frac{1-\alpha}{2^{\alpha}}\sum_{k=0}^{L-1}\int_{k/L}^{(k+1)/L}\frac{\cos q(y)}{\Vert x-y\Vert^{\alpha}}\: dy,\end{equation}
using the regularity of $1/\Vert x-y\Vert^{\alpha}$, we can extract
it from the integral:\begin{equation} 
C(x)\approx\frac{1-\alpha}{2^{\alpha}}\sum_{k=0}^{L-1}\frac{1}{\Vert x-y_{k}\Vert^{\alpha}}\int_{k/L}^{(k+1)/L}\cos q(y)\: dy, \label{C_approx} \end{equation}
with $y_{k}\in[k/L,\!(k+1)/L]$ %
The above approximation can be rigorously derived via a detailed
expansion that we enclose in the annexed appendix.
Taking advantage from the latter expression and aiming at reproducing a non trivial state with $C(x)=M$, we 
now assign the $q$ values on a circle so that in any small interval
the average of $\cos q(x)$ is constant and equal to the global, constant magnetization $M$
of the system. This procedure implies a peculiar spatial organization
which returns a constant coarse grained image of the function $\:\cos q$,
equal in turn to $M$. Note also that a similar coarse grained procedure is adopted in  \cite{Barre05}, with reference to a different case study. 
In doing so, and as a straightforwad consequence of Eq.(\ref{C_approx}), which ultimately holds because of the continuum limit hypothesis, we build up
a configuration which returns the sought condition: $C(x)$ is a constant function on the circle and equal to the magnetization $M$. 
The complex spatial organization of $(p(x),q(x))$ yielding to the constant coarse grained value of the magnetization $M$, can be in principle 
condensed into a local (i.e. dependent on $x$) one particle distribution function $f(p,q,x)$. Clearly, by construction, any  configuration $(p(x),q(x))$ compatible with $f(p,q,x)$, is going to be a solution of the fractional Eq.(\ref{eq:fractional_Steady_state})\footnote{Note that $x$ corresponds to a point on the circle for $(p(x),q(x))$, while, for for $f(p,q,x)$,  it labels the position $x$ of an interval of width  
$dx$.}  
However if we aim at building a configuration that shares some similarities with the aforementioned QSS, we still need stationarity, 
in the infinite N limit.
This property is guaranteed if we refer to the stationary distributions that were calculated
for the HMF in \cite{Leoncini09b}. The recipe goes as follows: we pick up $q$ and $p$
values as originating from this stationary distribution, so as to ensure that
the time evolution will only consist of a local reshuffling of the
actual phase space coordinates, without affecting their associated
coarse grained value. In doing so, we hence obtain a family of stationary
solutions of the $\alpha$-HMF model. %
\footnote{We note that several distinct stationary distributions of the original
HMF model are in principle allowed, yielding to identical values of
$M$. It should be hence possible to build up more complex distributions than those here analyzed, via a combination of 
such distinct profiles assigned to different spatial locations.
A stationary distribution of the
$\alpha$-HMF model can thus correspond to a stationary distribution
of the HMF model distributed scale free on the circle. Alternatively, it can be decomposed
in a sum of different stationary distributions, which all have
the same magnetization and are placed in different
regions of the circle. Notice that certain restrictions due to Eq.(\ref{eq:fractional_Steady_state}), may be necessarily imposed at the frontiers.%
}

The microscopic dynamics of the particles in such states can be mapped
into a pendulum motion and is hence integrable. Interestingly, the
spatial organization is locally scale free. The functions $q(x,t)$
and $p(x,t)$ are thus {}``very complicated'' along the spatial
direction, while displaying a regular time evolution and no chaos.
In order to validate this result for a finite size sample on the original
lattice %
\begin{figure}
\begin{centering}
\includegraphics[width=7cm]{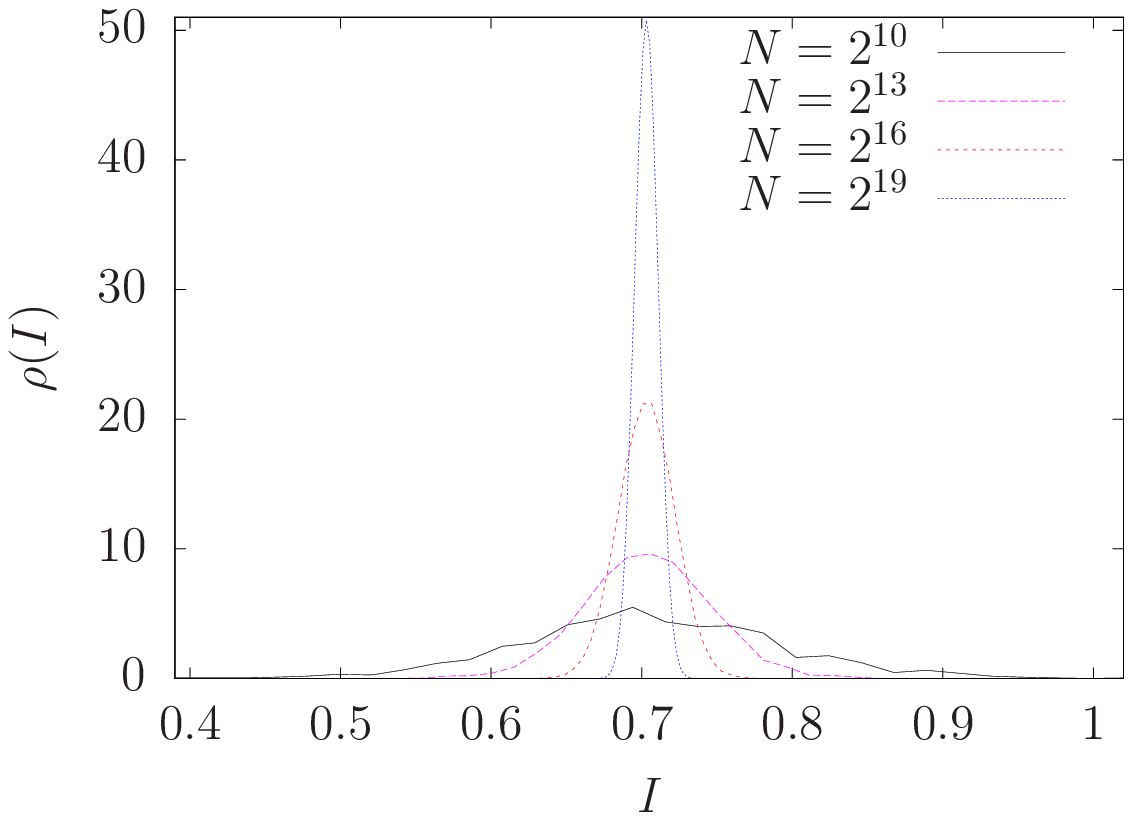}

\includegraphics[width=8cm]{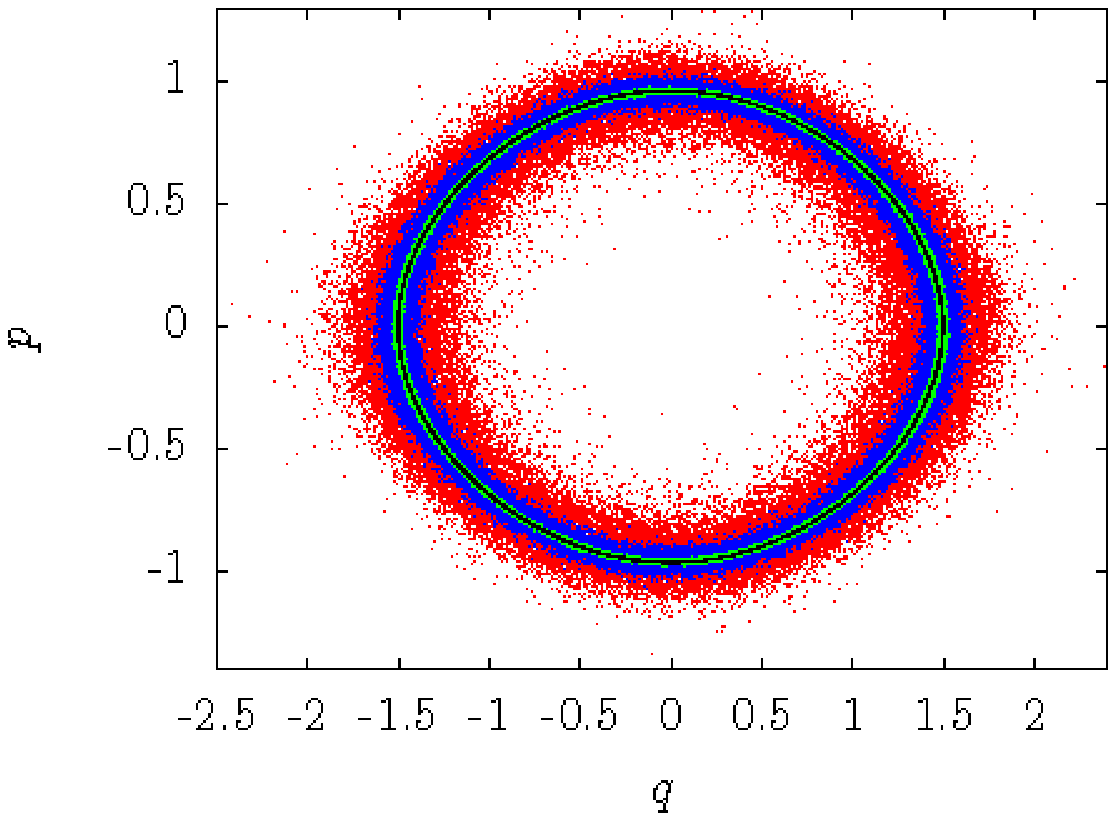}
\par\end{centering}

\caption{Top: QSS for the $\alpha$-HMF model built using a stationary solution
of the HMF model corresponding to a one particle PDF of the type $f(I,\theta)=\delta(I-I_{0})/2\pi$
(one torus $M=0.5$). The system is evolved via the $\alpha$-HMF
dynamics with $\alpha=0.25$. The figure displays the resulting distribution
of actions $I$ at $t=200$, for different sizes. The integration
uses the fifth order optimal symplectic integrator \cite{McLachlan92}
and a time step $\delta t=0.05$.  Bottom: A simple superposition
of snapshots relative to different values of $\alpha$ and at final time $t=200$,  with $N=2^{19}$. 
$\alpha=0.75$, $\alpha=0.5$, $\alpha=0.25$, $\alpha=0$, correspond
respectively to the (online) colors red, blue green and black. Clearly the thickness of the rings get reduced as 
$\alpha \rightarrow 0$, i.e. when the  $\alpha$-HMF model tends towards its corresponding HMF limit.  \label{fig:QSS-example 1}}

\end{figure}
 we proceed as follows. We consider a stationary state of the HMF
consisting of just one torus with associated magnetization $M=0.5$.
In pendulum action-angle variable we refer hence to a one particle
distribution of the type $f(I,\theta)=\delta(I-I_{0})/2\pi$ (see for instance
\cite{ZaslavBook98} for details regarding the transformation from $(p,q)$ to $(I,\theta)$ and vice versa).
 We are focusing our attention on an {}``individual component'' belonging
to the extended set of a linearly independent elements, which define
the QSS basis. In order to be as close as possible to the stationary
state of the $\alpha$-HMF model, we simply distribute randomly on
the lattice the values picked from such, analytically accessible,
distribution. The analysis for different values of the number of sites
and $\alpha=0.25$, is then performed by monitoring the values of
the action and shows as expected a trend towards the stabilization
of such a state as $N$ increases (see Fig.~\ref{fig:QSS-example 1}).
We also checked that as expected increasing the value of $\alpha$,
which weakens the coupling strength, implies inducing a more pronounced
destabilization of the state, which can be effectively opposed by
increasing the number of simulated rotors. The solutions here constructed
are hence stable versus the $\alpha$-HMF dynamics, provided the continuum
limit is being performed and so represent a consistent analytical
prediction for the existence of quasi- stationary states, beyond the
original HMF setting. 

\begin{figure}
\begin{centering}
\includegraphics[width=7cm]{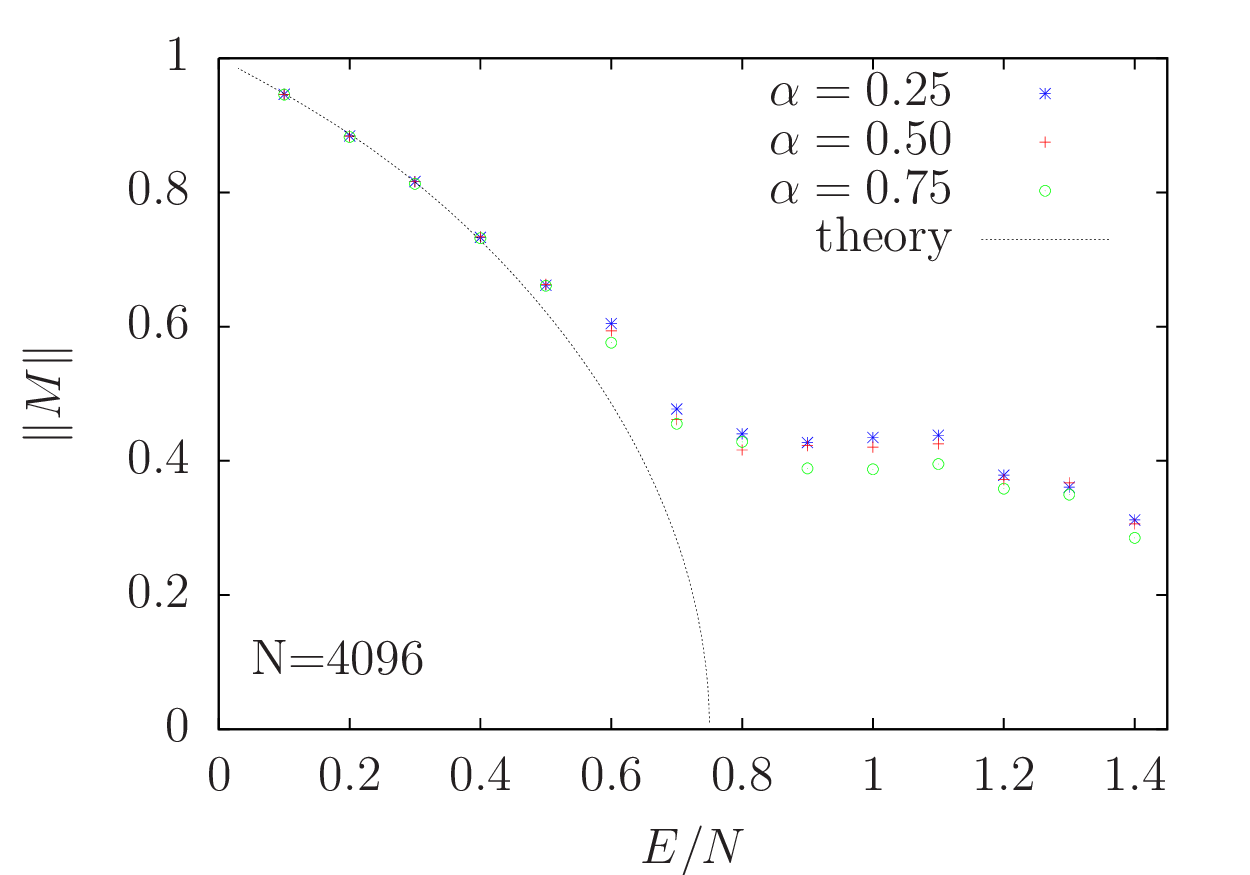}
\par\end{centering}

\caption{Magnetization vs. energy per particle $E/N$ for different $\alpha$,
for a system made of $N=4096$ particles. The magnetization values
here reported follow from a time average over the window $400<T<800$.
\label{fig:Energy-versus-magnetisation}}

\end{figure}

\begin{figure}
\begin{centering}
\includegraphics[width=7cm]{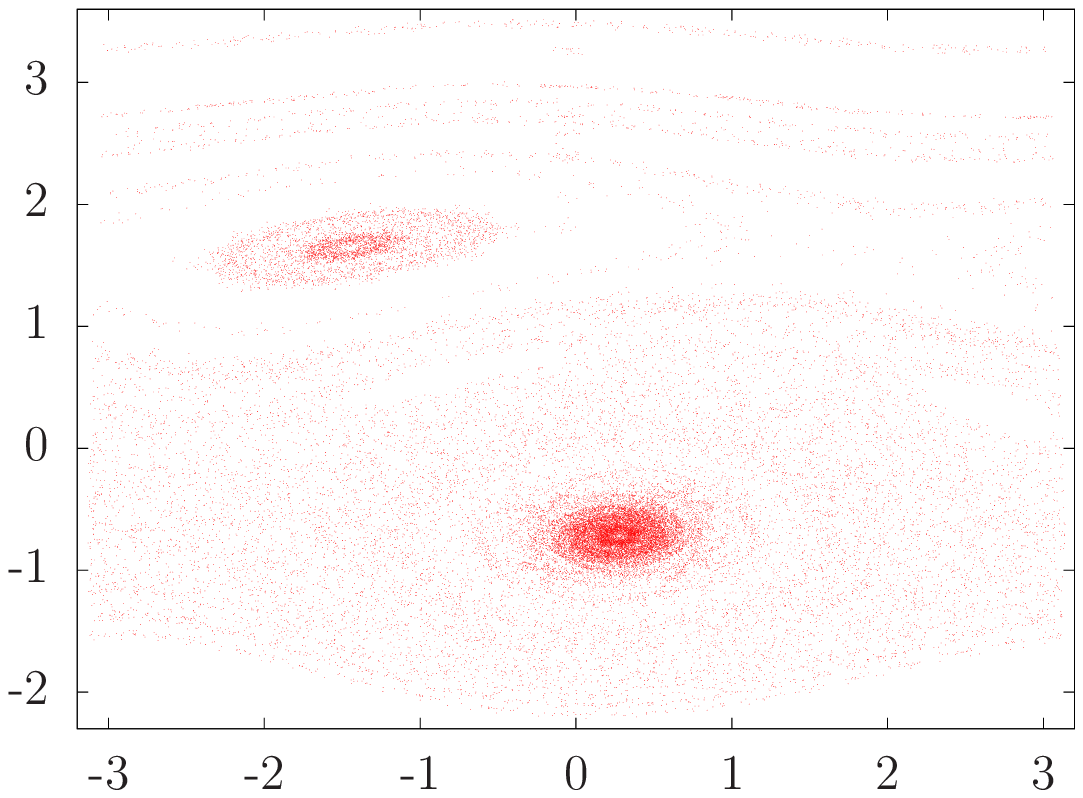}\\
\includegraphics[width=7cm]{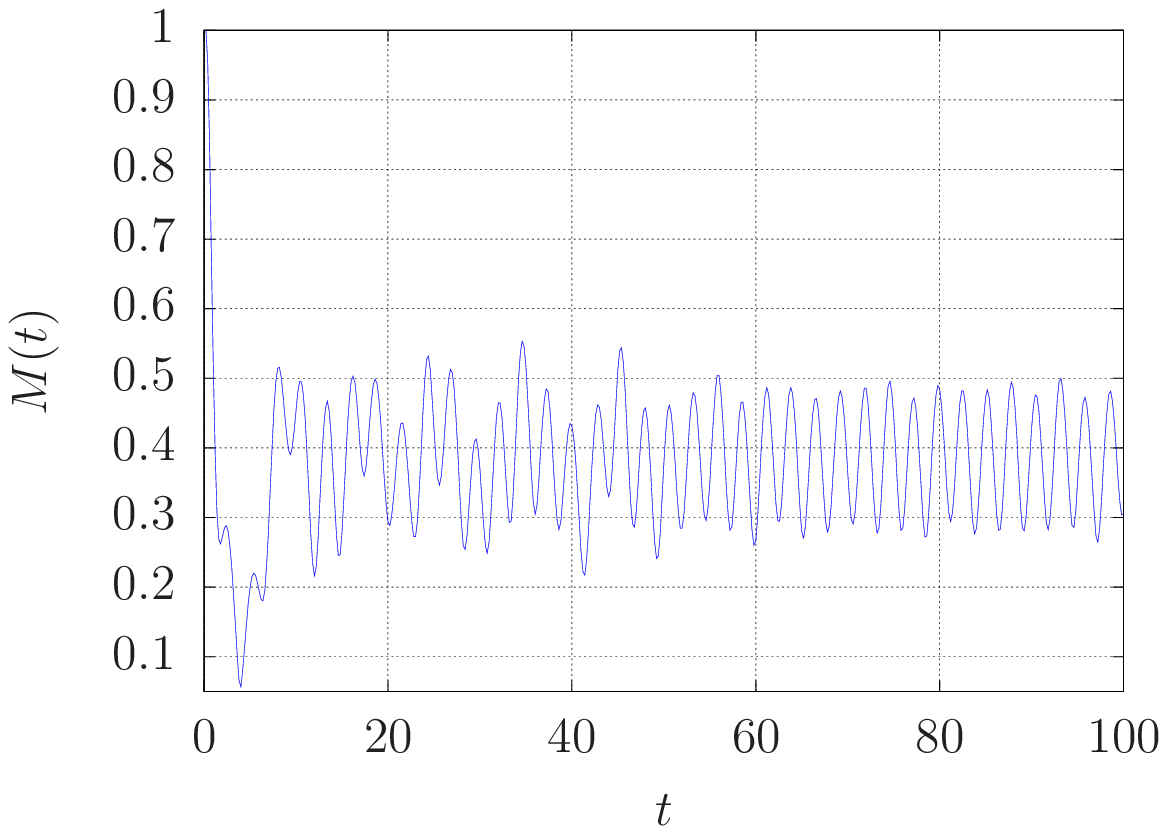}
\par\end{centering}

\caption{Top:{}``Poincar\'e section'' of a QSS for $E/N=1.2$, $N=65536$.
Initial conditions are Gaussian in $p$ and $q=0$. The section is
computed for $150<t<1000$ and $\alpha=0.25$. Bottom: Magnetization
versus time. \label{fig:Poincar=0000E9-section-of}}

\end{figure}

To further scrutinize the dynamics of the $\alpha$-HMF model we turn
to direct simulations, starting from out of equilibrium initial condition. Our declared goal is to follow the system evolution through 
the violent relaxation process and eventually identify the presence of spontaneoulsy emerging QSS for the $\alpha$-HMF model. 
In particular, we are interested  in their microscopic characteristic to make a bridge with the analysis developed in the first part of the paper.
For this purpose we initialize the system in $q=0$ and assume a Gaussian
distribution for the conjugate momenta $p$ \cite{Tamarit00}. The
system state is monitored by estimating the average magnetization
amount as a function of the energy, see Fig.~\ref{fig:Energy-versus-magnetisation}.
For energies larger than $0.75$ one would expect the homogeneous
solution to prevail, as dictated by the statistical mechanics calculation.
However, the system gets confined into a inhomogeneous state, the
residual time averaged magnetization being large and persistent in
time. It is therefore tempting to interpret those states as QSS, and
so analyze their associated dynamical features in light of the above
conclusions. In particular, we expect the microscopic dynamics to
resemble that of a pendulum, bearing some degree of intrinsic regularity.
To unravel the phase space characteristics we compute the Poincar\'e
sections, following the recipe in \cite{Bachelard08,Bachelard07}
and so visualizing the single particle stroboscopic dynamics, with
a rate of acquisition imposed by the self-consistent mean field evolution.
The averages of $C_{i}$ and $S_{i}$ refer to the two components
of the magnetization per site. The Poincar\'e sections are drawn by
recording the positions $p_{i}$ and $q_{i}-\varphi_{i}$ in phase
space each time the equality $M_{i}=M$ is verified. Results for a
specific initial conditions are depicted in Fig.~\ref{fig:Poincar=0000E9-section-of}
where one hundred trajectories are retained. %
 The phase portrait shares many similarities with that obtained for
a simple one and a half degree of freedom Hamiltonian (see for instance
\cite{ZaslavBook98}), with many resonances and invariant tori. Clearly,
and in agreement with the above scenario, a large number of particles
exhibit regular dynamics. However as the nature of phase space reveals,
these QSSs are steady states of the discrete dynamics, not stationary
solutions. %
\begin{figure}
\begin{centering}
\includegraphics[width=6cm]{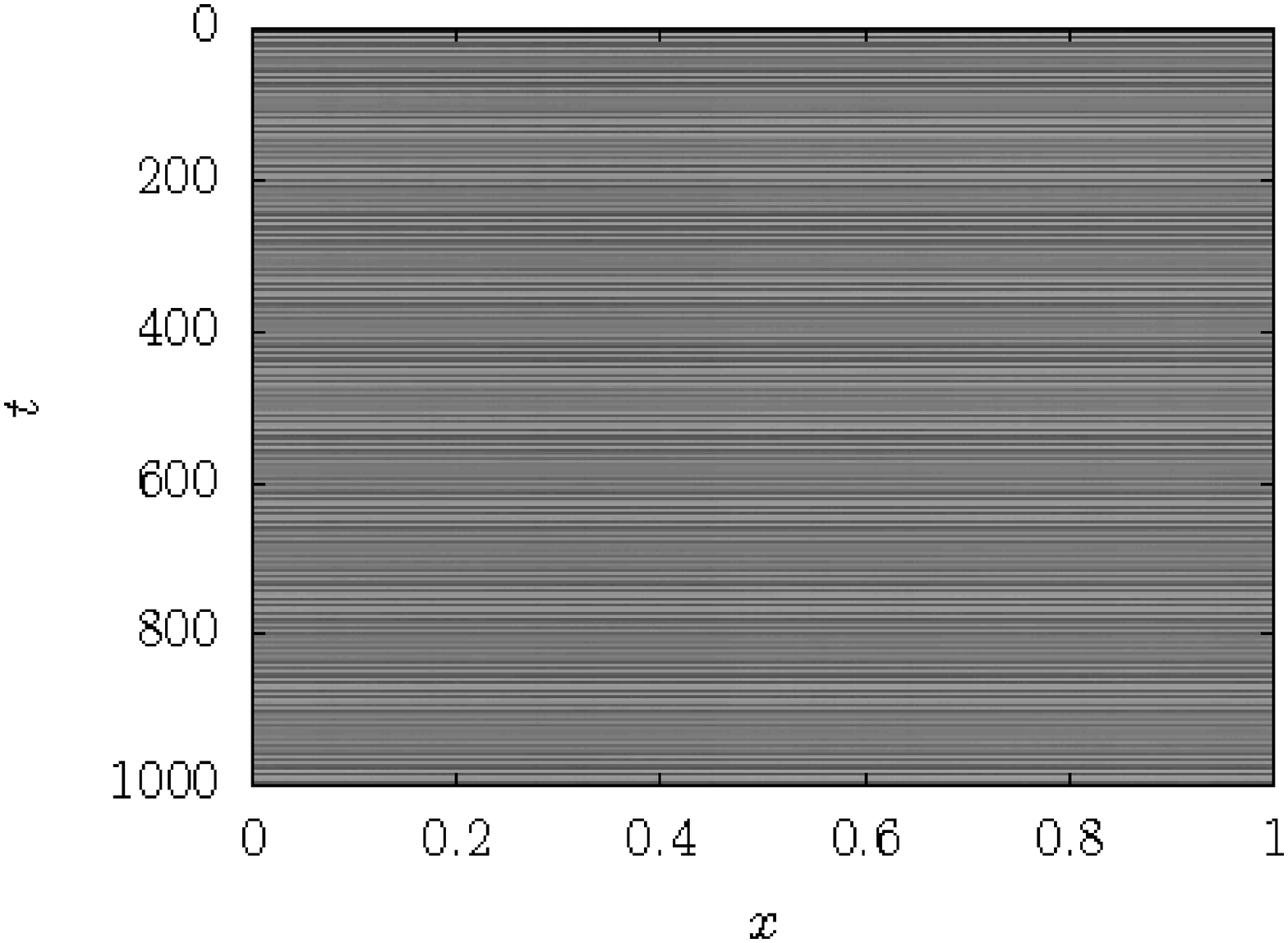}\\
\includegraphics[width=6cm]{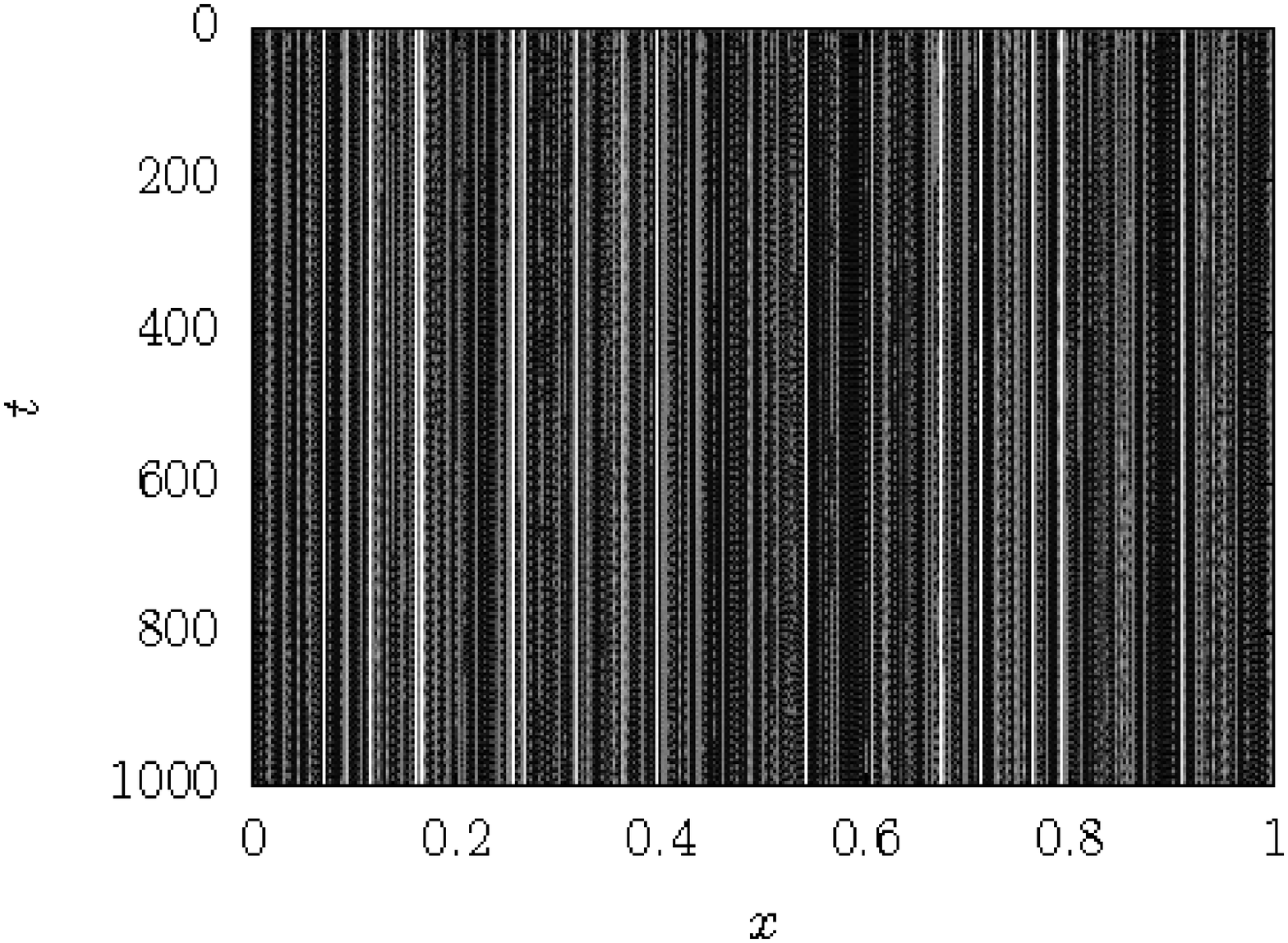}
\par\end{centering}

\caption{Top: $M(x)=\sqrt{C(x)^{2}+S(x)^{2}}$ as a function of time ,see also Fig.~\ref{fig:Poincar=0000E9-section-of} to monitor to appreciate 
the amplitude of the fluctuations as recorded along the time evolution. Bottom:
individual particle's action as a function of time. The system is
initialized to get a QSS with $E/N=1.2$ and $\alpha=0.25$ (see Fig.~\ref{fig:Poincar=0000E9-section-of}).
Total number of particle is $N=8192$, $x=i/N$, with $1\le i\le N$.
The actions are those of a pendulum corresponding to Eq.(\ref{eq:p_dot_bis}).
$M(x)$ is almost uniform in space, while the individual spatial organization
is complex. The time evolution of the actions appears quite regular.\label{fig:Spatial_Chaos}}

\end{figure}
 Nevertheless, we set to analyze the spatial organization of the identified
steady state to test whether stationary state features are present
in this configuration. To this end, we computed the values of the
local magnetization $M(x,t)=\sqrt{C(x,t)^{2}+S(x,t)^{2}}$ versus
time and estimated an individual action, stemming from a Hamiltonian
pendulum, which would give rise to an equation of motion formally
identical to the Eq.(\ref{eq:p_dot_bis}). Results of the analysis
are depicted in Fig.~\ref{fig:Spatial_Chaos}. One clearly sees that
the function $M(x,t)$ is homogeneous and presents a dependence
on $t$, thus suggesting that the distribution of $q(x,t)$ is a solution
of Eq.(\ref{eq:fractional_Steady_state}). The plot of action versus
time as depicted in Fig.~\ref{fig:Spatial_Chaos}, clearly indicates
a degree of enhanced spatial complexity, nearby particles not belonging
to the same tori. We find in these simulations and in this ($N$-
finite) steady state the same distinctive features of the stationary
solutions as depicted earlier. 

To conclude, in this Letter we have shown that stationary solutions
for the mean field HMF model are as well stationary ones of the $\alpha$-HMF
model. Microscopic dynamics in these stationary states is regular,
and explicitly known. The price to pay for this microscopic regularity
in time is a complex, self similar, spatial organization corresponding
to the solution of a fractional equation. When turning to direct numerical
investigations we have identified a series of quasi-stationary states,
which corresponds to steady states.
 Still, such states share many
of the features of their stationary counterparts. The importance of
these conclusions are manifold: On the one side, we confirm that QSS
do exist in a generalized non mean field setting \cite{Campa02}.
Also, we validate a theoretical method to construct, from first principle,
(out of) equilibrium stationary solutions. Finally, the fact that
in long range systems stationary states (among which we may of course
count the equilibrium) which display regular microscopic dynamics
do exist, allows us to dispose of an enormous amount of information
regarding the intimate dynamics of a system frozen in such state.
This knowledge can prove crucial in bridging the gap between  
microscopic and macroscopic realms in systems with many body interacting elements. Interestingly,  it could prove useful to investigate the role of chaos versus the foundation of statistical mechanics, as  
discussed in \cite{ZaslavBook98,Zaslav99}. 
\acknowledgments
We are very grateful to S. Ruffo for introducing us to this problem and for
providing useful comments and remarks. X. L. would also like to thank
F. Bouchet for fruitfull and inspiring discussions. D. F. thanks the 
EURATOM mobility program for financial support. 

\section{Appendix}

As mentioned in the core of the letter, we shall split the integral (\ref{eq:C_of_x})
in $L$ pieces. In this appendix we will justify the procedure that eventually leads to eq.( \ref{C_approx}).
 To this end and to make the derivation more transparent we will focus on $C(0)$ and
localize the  singularity in $0$. The generalization  
for $C(x)$ is straightforward. Let us start by defining
the small parameter $\epsilon=1/L$, and write: 
\begin{equation}
C(0)=\frac{1-\alpha}{2^{\alpha}}\left[\int_{0}^{\epsilon}\frac{\cos q(y)}{\Vert y\Vert^{\alpha}}+\sum_{k=1}^{L-1}\int_{k/L}^{(k+1)/L}\frac{\cos q(y)}{\Vert y\Vert^{\alpha}}\: dy\right]\;,
\end{equation}
Consider the intervals labelled with $k\ge1$, namely those 
where the singularity $\Vert y\Vert=0$ is not present. We
use the regularity of $1/\Vert y\Vert^{\alpha}$ and perform a Taylor
expansion of the function in the vicinity of $y_{k}=k/L$ to get
\begin{equation} 
\frac{1}{\Vert y\Vert^{\alpha}}=\frac{1}{\Vert y\Vert^{\alpha}}-\alpha\frac{y-y_{k}}{\Vert y_{k}\Vert^{\alpha+1}}+\dots
\end{equation}
which implies \begin{eqnarray*}
A_{k} & = & \int_{k/L}^{(k+1)/L}\frac{\cos q(y)}{\Vert y\Vert^{\alpha}}\: dy\approx\frac{1}{\Vert y_{k}\Vert^{\alpha}}\int_{k/L}^{(k+1)/L}\cos q(y)dy\\
 & - & \frac{\alpha}{\Vert y_{k}\Vert^{\alpha+1}}\int_{k/L}^{(k+1)/L}(y-y_{k})\cos q(y)dy+\dots\:.\end{eqnarray*}

Assume now that 
the average $\frac{1}{L}\int_{k/L}^{(k+1)/L}\cos q(y)dy$
is constant and equal to $M$, no matter the value of $k$. Furthermore, such a property is assumed to hold
as $L$ gets larger and larger. Then since $\cos(y)$
is bounded by one, we can get the following estimate 
\begin{equation}
|A_{k}-\frac{1}{\Vert y_{k}\Vert^{\alpha}}\frac{M}{L}|\lesssim\frac{1}{\Vert y_{k}\Vert^{\alpha+1}}\frac{1}{L^{2}}\:.\end{equation}
which constraints the $A_k$ term. 
Exploiting again the fact that $\cos(y)$ is bound and recalling the definition of $C(0)$ we end up with:
\begin{eqnarray*}
|C(0)-\frac{1-\alpha}{2^{\alpha}}\frac{M}{L}\sum_{k=1}^{L-1}\frac{1}{\Vert y_{k}\Vert^{\alpha}}| & \lesssim & \epsilon^{1-\alpha}+\frac{1}{L^{2}}\sum_{k=1}^{L-1}\frac{1}{\Vert y_{k}\Vert^{\alpha+1}}\\
 & \lesssim & \epsilon^{1-\alpha}+2\epsilon\int_{\epsilon}^{1/2}\frac{dy}{y^{\alpha+1}}\\
 & \lesssim & \epsilon^{1-\alpha}+\epsilon^{1-\alpha}\:.\end{eqnarray*}
This relation confirms the adequacy of the approximated expression (\ref{C_approx}), as it immediately follows by the definition of 
$M$ mentioned above. Furthermore, if we take the limit $\epsilon=1/L\rightarrow0$, we get  $C(0) \rightarrow M$, a result which is clearly true as long as $\alpha<1$.

\bibliographystyle{eplbib}

\end{document}